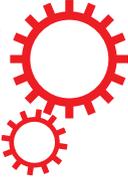

# Microenvironmental cooperation promotes early spread and bistability of a Warburg-like phenotype

Jorge Fernandez-de-Cossio-Diaz[1], Andrea De Martino[2,3] & Roberto Mulet[4]

We introduce an *in silico* model for the initial spread of an aberrant phenotype with Warburg-like overflow metabolism within a healthy homeostatic tissue in contact with a nutrient reservoir (the blood), aimed at characterizing the role of the microenvironment for aberrant growth. Accounting for cellular metabolic activity, competition for nutrients, spatial diffusion and their feedbacks on aberrant replication and death rates, we obtain a phase portrait where distinct asymptotic whole-tissue states are found upon varying the tissue-blood turnover rate and the level of blood-borne primary nutrient. Over a broad range of parameters, the spreading dynamics is bistable as random fluctuations can impact the final state of the tissue. Such a behaviour turns out to be linked to the re-cycling of overflow products by non-aberrant cells. Quantitative insight on the overall emerging picture is provided by a spatially homogeneous version of the model.

Overflow metabolism, namely the incomplete oxidation of nutrients at high intake rates leading to the secretion of partially metabolized products, is ubiquitous in cellular systems. Fast growth or high rates of energy expenditure are accompanied by the excretion of carbon equivalents like acetate in *E. coli*[1], ethanol in yeast[2], lactic acid in skeletal muscle cells[3], and lactate in many types of tumors[4] (where this is known as the Warburg effect). A similar scenario has been observed in immune cells[5]. Many studies have addressed the functional role[6] as well as the origin of this kind of metabolic re-programming at different levels, focusing most recently on the constraints on cellular energy metabolism imposed by different metabolic demands[7–9], macromolecular crowding[10], costs associated to gene expression[11–13], membrane occupancy[14] or intrinsic limits to mitochondrial oxidative phosphorylation[15], as well as on hypoxia[16] and dynamical mechanisms like flux sensors[17] (see refs [18], [19] for recent reviews of different models and possible mechanisms).

The effects of overflow in cell populations, however, specifically for the creation of functionally relevant microenvironments, are less studied. While it has been argued that the competition between slow-growing oxidizers and fast-growing fermenters may have played a central role in the evolution of multicellularity[20, 21], empirical evidence suggests that an effective cooperation between these distinct phenotypes may arise from the competition for nutrients, with important physiological consequences. The simplest manifestation of such a regime is via the shuttling of overflow products[22]. In short, fast growing cells typically outcompete their quiescent neighbours for the primary carbon source. In such conditions, the latter may have to re-adjust their metabolism, ultimately relying on the carbon equivalents excreted by the former for survival. This kind of cell-to-cell coupling has been found to play an important role in multicellular contexts ranging from cancer[23] to muscle cells[24] to brain energy metabolism[25, 26]. In cancer, the lactate-shuttle scenario has been analyzed theoretically[27] and *in silico*[28], experimentally through indirect genetic evidence[23], and has also been discussed as a potential therapeutic target[29] in the recent revival of interest in the re-programming of metabolism in cancer[30] (but see also ref. [31]). Taken together with the growing literature on tumor-stroma interactions[32, 33], these results present cancer as a strongly non-cell-autonomous disease and suggest a major functional role for the cooperative coupling between overflowing and non-overflowing cells. Similar scenarios have been studied in the context of microbial populations[34–36].

[1]Department of Systems Biology, Center of Molecular Immunology, La Habana, Cuba. [2]Soft and Living Matter Lab, Istituto di Nanotecnologia (CNR-NANOTEC), Rome, Italy. [3]Human Genetics Foundation, Turin, Italy. [4]Group of Complex Systems and Statistical Physics, Department of Theoretical Physics, Physics Faculty, University of Havana, La Habana, Cuba. Andrea De Martino and Roberto Mulet contributed equally to this work. Correspondence and requests for materials should be addressed to A.D. (email: andrea.demartino@roma1.infn.it)

Besides strictly energetic considerations related to the usefulness of overflow products (like lactate, acetate, ethanol and ammonia) as metabolic fuels, their shuttling might serve the added purpose of clearing the extracellular space of the potentially toxic side-effects caused by their accumulation. The toxicity of lactate and ammonia has been well characterized for mammalian cell cultures[37, 38] and cancer[39, 40], in which case the increased microenvironment acidity caused by extracellular lactate has been found to stimulate the death of surrounding cells[41, 42]. In turn, acetate can inhibit *E. coli* growth by as much as 50%[1, 43] while yeast's tolerance to ethanol is known to be strongly strain-dependent[44]. Likewise, industrial mammalian cell cultures are commonly affected by the accumulation of toxic metabolites secreted during the production process and exhibit qualitatively distinct metabolic states in response to variations in extracellular metabolite concentrations, transitioning from efficient, oxidative phenotypes to inefficient wasteful states[45]. In this context the accumulation of waste is usually undesired and one strives to maintain cells that can fully metabolize their nutrients into compounds of interest, posing an important biotechnological challenge[46].

Here we attempt to study *in silico* how carbon overflow affects the overall metabolic organization of cells in a tissue, including spatial effects as well as effects due to shuttling, microenvironment toxification and exchanges with the blood. Our scenario is close to that considered in models of avascular tumor growth[47–49] but integrates a coarse-grained perspective of cell metabolism along the lines of refs [28, 50] and the acidification of the environment due to the byproducts of metabolism. In particular, we assume that, in a tissue formed by healthy cells which can exchange compounds with a chemical reservoir, a single aberrant cell is planted, with upregulated primary nutrient transporters and disregulated energetics leading to carbon overflow and (stochastically) aberrant cell death and/or replication. We analyze the conditions under which the competition for the primary carbon source can be turned into a cooperative regime where non-aberrant cells are sustained by the aberrant ones' overflow product, in turn fostering the spread of the latter. By extensive numerical simulations, we obtain a "phase structure" in which the outcome is studied as a function of the concentration $c$ of primary nutrient in the reservoir and of the characteristic turn-over rate $k$ for tissue-reservoir exchanges. We furthermore develop a mathematically solvable version of the model where spatial effects are neglected, allowing to unveil the physical origin of many of the key features found in its spatial counterpart. Our results suggest that the accumulation of waste can limit the growth of the aberrant cells, leading to their demise, unless remediated either by re-cycling via healthy cells or by export away from the tissue. In such a scenario, $c$ and $k$ are the key variables controlling whether the aberrant cells will spread to the entire tissue. However, in the spatial model, random fluctuations can drive the tissue toward different outcomes. This bistability is absent in the spatially homogeneous version and is tightly coupled to the shuttling of overflow products.

## Results

**Model definition and properties.** *Basic setup.* We consider an ensemble of cells (a "tissue" for short) in contact with a nutrient reservoir (the "blood", see Fig. 1A). The tissue is modeled as a single layer of cells arranged in an $N \times N$ square lattice, with each site $i$ occupied by a cell. A primary carbon source $S$ (say, glucose) is available in the blood at a fixed concentration denoted by $c$, and can diffuse through the blood-tissue barrier and across the tissue. Likewise, a waste product of the metabolism of cells within the tissue can diffuse across the tissue or be exported into the blood. The effect of metabolite exchange with the blood is quantified by an effective metabolite turn-over time $\tau$.

*Cellular metabolic networks.* Cells can be of two types, which we shall call "healthy" and "aberrant" respectively. The two types differ slightly in their metabolic capabilities. Specifically, each cell $i$ can import and process $S$ at a rate $u_i$ that depends on the local substrate level $s_i$ and on its energy demand (see Fig. 1B). The in-taken nutrient is processed into a partially metabolized precursor $P$ and energy $E$ (i.e. ATP), with a theoretical energy yield equal to $Y_u$ (i.e. at most $Y_u$ units of energy are obtained per unit of $u_i$). However, while aberrant cells can turn $P$ into a 'waste' metabolite $W$ and excrete it, healthy cells can import $W$ and re-cycle it to produce $P$. The corresponding fluxes $v_i$ ($v_i \geq 0$ for importing healthy cells, $v_i \leq 0$ for exporting aberrant cells) are determined by the local level of $W$ in the tissue (denoted by $w_i$) and by cellular energy demands. Finally, energy can be derived from $P$ at rate $r_i$ given, by mass-balance, by $r_i = u_i + v_i \geq 0$. Assuming a yield for the $P \to E$ conversion equal to $Y_r$, the energy output flux obtainable from $S$ and $W$ is

$$e_i = Y_u u_i + Y_r r_i \equiv Y_u u_i + Y_r (u_i + v_i). \tag{1}$$

Identifying $r_i$ with the flux through respiration, we assume it is capped[50] so that $0 \leq r_i \leq r_{\max}$. We furthermore assume that $Y_u < Y_r$, by analogy with the universally lower yield of fermentation versus respiration.

*Metabolic objectives and optimal flux patterns.* Aberrant and healthy cells also differ strongly both in their maximum $S$-intake capability, with aberrant cells assumed to be able to sustain much higher values of $u_i$ than healthy cells, and in their respective metabolic goals, in that the former are assumed to maximize their energy output (1) while the latter only aim at generating a fixed maintenance energy output flux $e_m$. In addition, we conform to empirical *in vitro* studies showing that cells may consume a waste product of their metabolism like lactate or acetate only after glucose is depleted[46, 51], and assume that cells employ the primary nutrient preferentially. These assumptions lead to specific forms for the key fluxes $u_i$ and $v_i$. Defining $u_{MM}(s_i) = u_{\max} s_i / (K_s + s_i)$ and $v_{MM}(w_i) = v_{\max} w_i / (K_w + w_i)$ as the fluxes of standard Michaelis-Menten uptake kinetics with parameters $(u_{\max}, K_s)$ and $(v_{\max}, K_w)$, and denoting by $u_m \equiv e_m / (Y_u + Y_r)$ the smallest value of $u_i$ ensuring that $e_i = e_m$ without resorting to $W$ (i.e. for $v_i = 0$) one has that, for healthy cells,

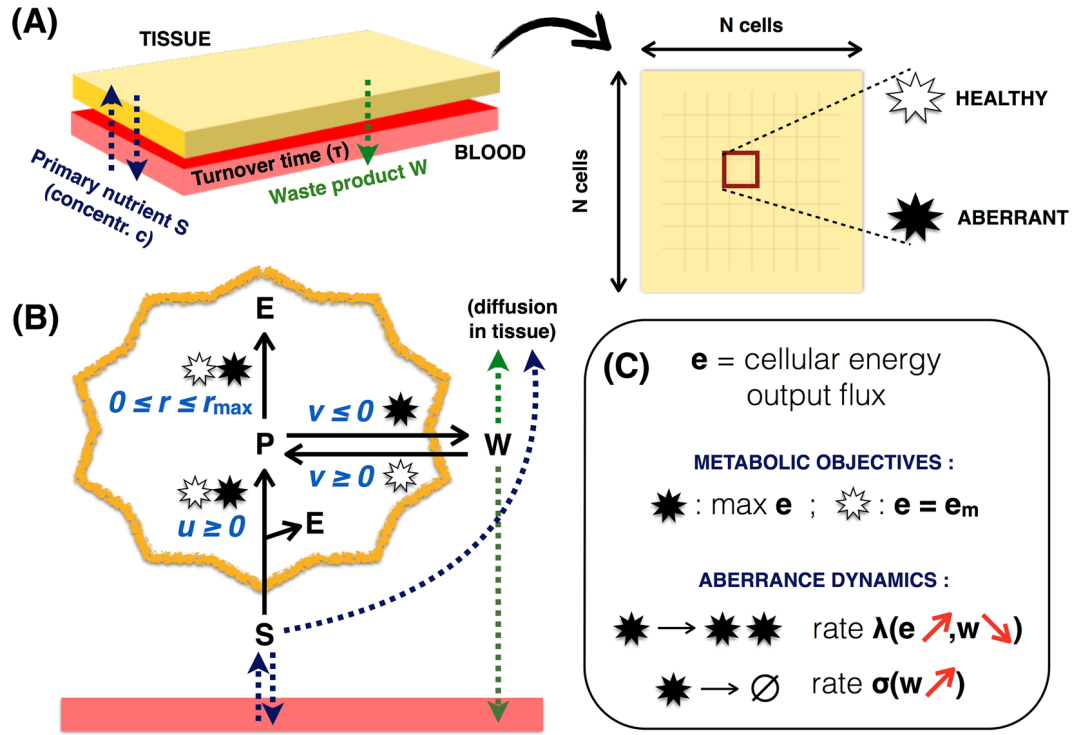

**Figure 1.** Scheme of main processes considered in the model. (**A**) A single layer of cells arranged in an $N \times N$ square lattice (the "tissue") is in contact with a reservoir (the "blood"), where a primary carbon source $S$ is available at concentration $c$. $S$ can diffuse through the blood-tissue barrier and inside the tissue. Each tissue site is occupied by a cell that can be either "healthy" or "aberrant". (**B**) Each cell can import and process $S$ generating energy $E$ (rate $u$, yield $Y_u$) and a precursor $P$ and further process the precursor into $E$ (rate $r$, yield $Y_r > Y_u$). Because $r$ is capped, aberrant cells can process the excess precursor into a waste product $W$, which is excreted and can diffuse across the tissue and into the blood. Healthy cells instead operate far from saturating $r$ and can re-cycle $W$ to generate more $P$. (**C**) Aberrant cells have a larger capacity to import $S$ and aim at maximizing their energy output flux $e$. $e_m$ instead represents the maintenance energy output of healthy cells. Cells are sensitive to the local concentrations of $W$ (denoted by $w$) and $S$ (denoted by $s$). In particular, aberrant cells replicate at a rate $\lambda$ that is an increasing function of their energy output and a decreasing function of $w$. Likewise, they can die (and be replaced by a healthy cell) at a rate $\sigma$ that is an increasing function of $w$. The primary nutrient import flux is instead a function of the local level of $S$ experienced by cells.

$$u_i = \min\{u_{\text{MM}}(s_i), u_m\} \qquad (2)$$

$$v_i = \min\left\{v_{\text{MM}}(w_i), (u_m - u_i)\left(1 + \frac{Y_u}{Y_r}\right)\right\} \qquad (3)$$

whereas for aberrant cells

$$u_i = \alpha_{\text{ab}} u_{\text{MM}}(s_i) \qquad (4)$$

$$v_i = \min\{0, r_{\max} - u_i\} \qquad (5)$$

with $\alpha_{\text{ab}} \gg 1$ a constant quantifying the dysregulated uptake capacity of aberrant cells. In words, for healthy cells, if $u_{\text{MM}}(s_i) > u_m$ then $u_i = u_m$ and no in-take of $W$ is required to ensure that $e_i = e_m$ (i.e. $v_i = 0$). On the other hand, for $u_{\text{MM}}(s_i) < u_m$ cells will import as much of $S$ as possible and then import either the minimal amount of $W$ required to meet their energy demand $\left(v_i = (u_m - u_i)\left(1 + \frac{Y_u}{Y_r}\right)\right)$ or as much $W$ as possible according to local availability ($v_i = v_{\text{MM}}(w_i)$). For aberrant cells, instead, if $u_i < r_{\max}$ then all the primary nutrient imported can be processed through $r_i$ and no precursor is excreted ($v_i = 0$). When $u_i$ exceeds $r_{\max}$, though, $W$ is produced and secreted at net rate $|v_i| = u_i - r_{\max}$.

*Aberrant cells death and replication dynamics.* A feedback between the population dynamics of aberrant cells and waste accumulation in the tissue is established by the fact that both their replication ($\lambda_i$) and death ($\sigma_i$) rates depend on the local level of waste $w_i$ (see Fig. 1C). In particular, we assume that every aberrant cell with at least one healthy neighbor can replicate, with the daughter cell occupying the site of a randomly picked normal neighbor, with a probability per unit time given by

$$\lambda_i = \lambda_{\max} \frac{\max\{e_i - e_m, 0\}}{e_i - e_m + K_E} \frac{K_{\text{inh}}}{K_{\text{inh}} + w_i}, \quad (6)$$

where $\lambda_{\max}$ represents the fastest growth rate achievable by aberrant cells while $K_{\text{inh}}$ and $K_E$ are constants. In short, the growth rate is zero unless cells exceed the maintenance energy flux $e_m$, it increases in a Monod fashion as $e_i$ grows, but gets smaller as $w_i$ increases[52]. The particular choice for the dependence on $w_i$ avoids the introduction of extra parameters (compared for instance to a Hill function) while returning the same qualitative behaviour. Likewise, we assume the death rate $\sigma_i$ to be given by refs [53], [54]

$$\sigma_i = \sigma_{\max} \frac{w_i}{K_{\text{tox}} + w_i}, \quad (7)$$

which becomes larger (and tends to the limiting value $\sigma_{\max}$) as $w_i$ increases. We assume that aberrant cells are replaced by healthy ones upon dying. Note that we don't consider a specific birth/death dynamics for healthy cells, implicitly assuming that they are found in all tissue sites that are not occupied by aberrant cells (e.g. both aberrant and healthy cells are immediately replaced by healthy ones at death). In such conditions, the growth dynamics of healthy cells is immaterial.

*Diffusion of primary nutrient and waste.* Finally, the local concentrations of the primary nutrient ($s_i$) and of the precursor ($w_i$) are assumed to change in time and space according to

$$\frac{ds_i}{dt} = D_S \left( \frac{1}{d} \sum_{j \in \mathcal{N}(i)} s_j - s_i \right) - u_i + k(c - s_i) \quad (8)$$

$$\frac{dw_i}{dt} = D_W \left( \frac{1}{d} \sum_{j \in \mathcal{N}(i)} w_j - w_i \right) - v_i - k w_i \quad (9)$$

where $k \equiv 1/\tau$. The first terms on the right-hand-sides of (8, 9) represent the local concentration smoothing due to diffusion, with $D_S$ and $D_W$ the diffusion coefficients, $d$ the number of neighbors of cell $i$ ($d = 4$ in our case) and the sum runs over the set $\mathcal{N}(i)$ of neighbors of cell $i$. The second terms describe the changes due to import or export of compounds by cells, with $u_i$ and $v_i$ depending on concentrations as discussed in (2), (3), (4) and (5). The third terms denote instead the variations due to the exchange of primary nutrient and, respectively, the removal of the precursor from the tissue, in which the blood acts as an external reservoir for $S$ and as a sink for $W$.

*Time evolution.* After initiating the system in a configuration with a single aberrant cell among healthy cells (time $t = 0$), the dynamics of aberrance spread depends on (i) the pattern of local availability of primary nutrient, and (ii) how efficiently growth-suppressing waste can be removed from the tissue, either via the blood or via recycling by healthy cells. Overall, this scenario describes the conditions for the emergence of a mutually beneficial (local) microenvironment, characterized by a $W$-mediated metabolic coupling between healthy and aberrant cells. We are interested in quantifying its role in the sustainment of aberrant growth and the survival of healthy cells, and in understanding how it relates to the macroscopic environmental parameters $c$ (the concentration of primary nutrient in the blood) and $\tau = k^{-1}$ (metabolite turn-over time).

**Dynamics of aberrant spread.** We carried out extensive numerical simulations of the model upon changing the key environmental parameters $c$ and $k$, after having rendered parameters non-dimensional and having set them at realistic representative values (see Methods). Typical snapshots from a full-scale simulation of a $20 \times 20$ tissue obtained for $k = 10^3$ and $c = 0.1$ are shown in Fig. 2A. Initially, the aberrance spreads and waste is secreted (red markers, $v < 0$), while nearby healthy cells are deprived of primary nutrient and start to rely on the waste secreted by aberrant cells for survival (pink markers, $v > 0$). Growth of the aberrant population is therefore sustained by the coupling with healthy cells. Eventually, though, the primary nutrient gets depleted and aberrant cells in the core no longer saturate the respiration flux (black markers, $v = 0$), thereby ceasing waste secretion.

Figure 2B,C summarize the main features of the time evolution of the system observed in the simulation. The mean local concentration of primary nutrient $s$ decreases with time while the average waste level $w$ increases rapidly until it peaks (black arrow). At that point, healthy cells begin to in-take and recycle the waste product of aberrant growth for survival when their maintenance demands cannot be met from the primary nutrient alone (see Fig. 2C). A noticeable acceleration in aberrant growth then occurs (viz. the change in slope at the black dashed line in Fig. 2B) as the waste concentration begins to decrease and the environment becomes less toxic. As $s$ decreases, $W$ becomes the main metabolic fuel of healthy cells, while a fraction of aberrant cells in the core stops secreting $W$, further pushing down the level of waste. At the peak of waste uptake by healthy cells (green arrow), a second inflection point in aberrant growth can be observed (viz. the slope change at the green dashed line in Fig. 2B). The mean waste level vanishes shortly thereafter. Eventually the entire tissue is taken over by aberrant cells and $\phi$ saturates at one.

**Phase portrait.** Results concerning dynamical features and final configurations can be summarized in the 'phase portrait' displayed in Fig. 3A in terms of the asymptotic value of the mean fraction $\phi$ of aberrant cells in the tissue as a function of $c$ and of the effective blood-tissue turn-over rate $k$ (averages over 20 runs). At low nutrient level and long turnover times (region labeled **W−**), the primary source cannot support aberrant growth and the tissue ultimately stays healthy. In this region, no waste accumulates in the tissue and therefore no waste recycling occurs. Upon increasing $k$ and/or $c$ (region labeled **Sh**.), one enters a phase in which metabolic coupling with waste recycling occurs during the dynamics. Interestingly, though, if $k$ gets sufficiently large, the aberrance can spread to the whole tissue, strongly implicating the metabolic coupling and/or enhanced remediation through the bloodstream as keys for tissue takeover by aberrant cells. A third regime (region labeled **W+**) occurs if $k$ or $c$ are sufficiently large, in which healthy cells can meet their energy demand thanks to the large availability of primary substrate, so that no recycling takes place. If the increase in $c$ is not coupled to an increase in $k$, waste accumulates in the tissue because blood exchange alone does not suffice to remove $W$ ($k$ too low), thereby repressing aberrant growth due to increasing toxicity. As a consequence, the tissue stays healthy in the long run. On the other hand, if $k$ also increases, aberrant cells ultimately take over the tissue as the growth-suppressing waste product is efficiently cleared from the tissue thanks to the high turn-over rate.

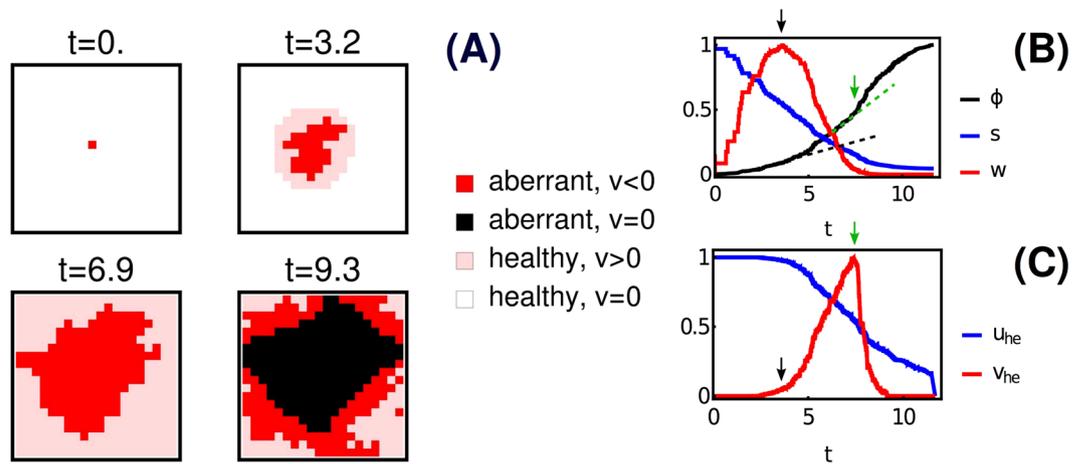

**Figure 2.** Simulation results for the case of fast turn-over rate (large $k$) and low blood nutrient level (small $c$). (**A**) Snapshots of the tissue. Aberrant growth is sustained by the waste-mediated coupling with healthy cells, which re-cycle $W$ to survive, thereby allowing for a fast replication rate of aberrant cells. Where no re-cycling occurs, aberrant cells no longer excrete $W$ but cease to grow. Time is in units of the inverse death rate (see Methods). (**B**) Fraction of aberrant cells ($\phi$), mean tissue level of primary nutrient ($s$) and mean tissue level of waste ($w$) as a function of time. (**C**) Mean primary nutrient intake flux ($u_{he}$) and mean waste intake flux ($v_{he}$) by healthy cells. The curves in (**B**,**C**) are normalized to a maximum of 1. The arrows in (**B**,**C**) mark significant points of inflection in the growth of aberrant cells, discussed in the text.

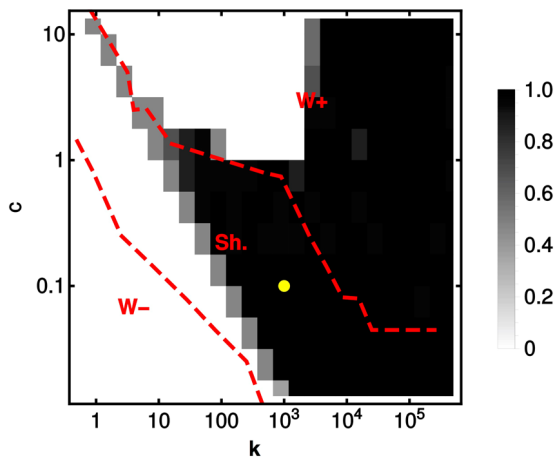

**Figure 3.** Phase portrait of the spatial model in the space spanned by the blood-tissue turn-over rate $k$ and nutrient availability $c$. Asymptotic value of the fraction $\phi$ of aberrant cells (color coded) as a function of the turn-over rate $k \equiv \tau^{-1}$ and of the level $c$ of primary nutrient in the blood. Dashed red lines delimit regions with qualitatively different roles for the waste. In region labeled W− no waste is secreted; in region Sh., waste secretion by aberrant cells occurs coupled with waste recycling by healthy cells; in region W+, waste secretion but no re-cycling takes place. Simulations are carried out in a $20 \times 20$ grid, and results are averaged over 20 different runs per point. The yellow dot marks the parameters used in Fig. 2.

**Insights from a solvable mathematical model.** *Spatially homogeneous model.* The origin of the rich behavior observed *in silico* can be mathematically traced by a simplified, analytically tractable version of the model. The simplification consists in assuming spatial homogeneity, i.e. that the transport and diffusion of the primary substrate *S* and of the waste product *W* are sufficiently fast for all cells to sense the same local levels of both species. This leads to a "mean-field" type of model that ignores the spatial organization of cells. The state of the tissue can be fully described in terms of three variables, namely the fraction of aberrant cells $\phi$, the level of primary nutrient in tissue *s* and the level of waste product in tissue *w*. Assuming that the daughters of aberrant cells replace normal cells in a population constrained to a fixed total size, the above quantities evolve in time according to

$$\frac{d\phi}{dt} = (\lambda - \sigma)\phi(1 - \phi) \tag{10}$$

$$\frac{ds}{dt} = -\phi u_{ab} - (1 - \phi)u_{he} + k(c - s) \tag{11}$$

$$\frac{dw}{dt} = -\phi v_{ab} - (1 - \phi)v_{he} - kw \tag{12}$$

where the parameters $\lambda$ (replication rate of aberrant cells), $\sigma$ (death rate of aberrant cells), $u_{he}$ and $u_{ab}$ (primary nutrient intake flux of a healthy and, respectively, aberrant cell), and $v_{he}$ and $v_{ab}$ (the waste intake and, respectively, out-take flux of a healthy and, respectively, aberrant cell) depend in turn on *s* and *w* as in Equations (2)–(7). Equation (10) describes the change in $\phi$ due to duplication and death events (both of which require the vicinity of a healthy cell) at fixed total number of cells. The evolution of *s* and *w*, instead, is due to cellular import and export activities and to exchanges with the blood.

Coupling fast homogenization with a quasi-steady-state assumption for concentration variables (based on the fact that the time scales characterizing cell replication are much longer than those over which tissue levels of primary nutrient and waste equilibrate) allows to decouple the dynamics of *s* and *w* from that of $\phi$. Specifically, setting $\frac{ds}{dt} = \frac{dw}{dt} = 0$, one may express both quantities as functions of $\phi$, thereby leaving the latter as the key variable of the model. Numerical solution of its equilibrium value from Eq. (10) (see Methods for the choice of parameters) finally results in the phase portrait displayed in Fig. 4A. The homogeneous model appears to match closely its spatial counterpart (Fig. 3) except for a key detail, provided in Fig. 4B. In specific, one sees that the spatial model displays coexistence of fully aberrant states with fully healthy states for low *k* and intermediate *c* (region Sh.), i.e. for the same values of *c* and *k* different runs may end up with the tissue being taken over by aberrant cells or with a completely healthy tissue, while the simplified model is unable to capture this aspect.

Cutting through the (*k*, *c*) diagram at different levels of *c* (displayed as yellow lines in Fig. 4A) different scenarios are highlighted (see Fig. 4C). For low *c*, $\phi$ gradually increases with *k* until it saturates close to the maximum. Upon further increasing *k*, one generates a waste shuttling from aberrant to healthy cells with accumulation of waste in the tissue, which is remediated at higher *k* by faster wash-away into the blood stream. At high *c*, instead, aberrant cells can survive even at low *k* without secreting waste. If *k* increases, though, waste excretion sets in that is not managed since (i) the local level of primary nutrient is large enough for healthy cells to survive based on it alone, and (ii) removal by the blood is still too slow. As a consequence, aberrant growth is repressed and $\phi$ decreases. A further increase of *k* however allows to re-start aberrant growth by clearing up the micro-environment.

*The state of the tissue is regulated by a series of thresholds.* It is remarkable that the transition between full aberrant dominance and healthy tissue in some cases turns out to be discontinuous, viz. the separation between the Sh. and W+ regions marked by a blue line in Fig. 4A and by a blue arrow in Fig. 4C. In such conditions, small variations in *k* or *c* suffice to cause large-scale changes in the final tissue state. The same behaviour is observed in the fully spatial model (cf. Fig. 4B). This highlights the existence of thresholds where the qualitative characteristics of aberrant growth and tissue metabolism change drastically. Within the solvable model, it is possible to fully trace their origin to the parameters of the model. Mathematical analysis (see Supporting Text) leads in particular to the identification of three thresholds for the tissue level of primary substrate, i.e.

$$s_m = \frac{K_s u_m}{u_{max} - u_m}, \quad s_{sec} = \frac{K_s r_{max}}{\alpha_{ab} u_{max} - r_{max}}, \quad s_{gr} = \frac{K_s u_m}{\alpha_{ab} u_{max} - u_m}, \tag{13}$$

or, equivalently, in terms of the fraction of aberrant cells (see Fig. 5A),

$$\phi_m = \frac{k(c - s_m) - u_m}{(\alpha_{ab} - 1)u_m}, \quad \phi_{sec} = \frac{\alpha_{ab} k(c - s_{sec}) - \min(r_{max}, \alpha_{ab} u_m)}{\alpha_{ab} r_{max} - \min(r_{max}, \alpha_{ab} u_m)}, \quad \phi_{gr} = \frac{\alpha_{ab} k(c - s_{gr}) - u_m}{(\alpha_{ab} - 1)u_m}, \tag{14}$$

using which the phase structure can be understood. Note that we assume $u_m \leq r_{max}$ (see Methods), so that $s_{gr} \leq s_{sec} \leq s_m$.

Starting from (13), healthy cells are found to be able to sustain their maintenance demands from the primary nutrient alone if $s \geq s_m$, while for $s < s_m$ they must rely on re-cycling of *W* (if available). Aberrant cells, on the other hand, can have a non-zero replication rate (Eq. 6) if $s > s_{gr}$ and they secrete *W* if $s > s_{sec}$. Therefore, a metabolic coupling with shuttling of *W* from aberrant to healthy cells may only exist in the range $s_{sec} < s < s_m$. (In turn, $s_m > s_{sec}$ or, equivalently, $r_{max} < \alpha_{ab} u_m$ is a necessary condition for metabolic coupling).

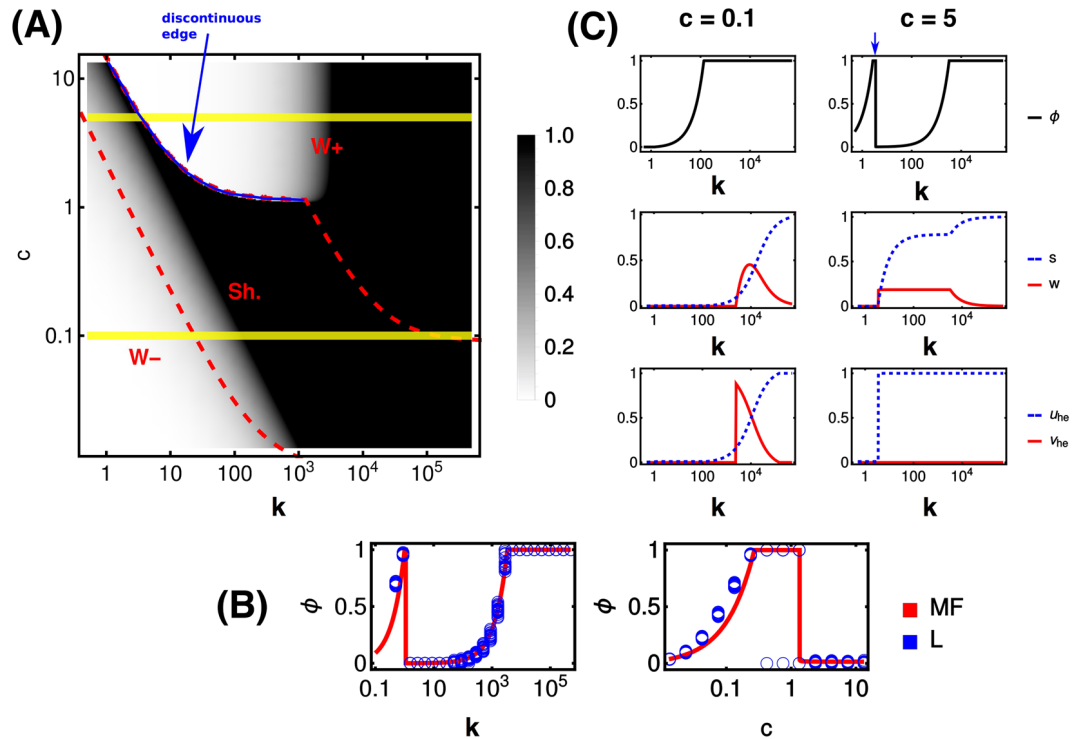

**Figure 4.** Summary of results for the spatially homogeneous model. (**A**) Fraction $\phi$ of aberrant cells (color-coded) as a function of the primary nutrient level in the blood ($c$) and of turn-over rate ($k$). Regions W−, Sh. and W+ are as in Fig. 3. The blue line marks the discontinuous transition from a healthy to a fully aberrant tissue at the upper separation between regions Sh. and W+. (**B**) Comparison between the fully spatial and mean-field models. We show $\phi$ versus $k$ at fixed $c=7.5$ and $\phi$ versus $c$ at fixed $k=50$ for the former (blue markers, obtained in many different runs) and the latter (red line). (**C**) Fraction of aberrant cells ($\phi$), tissue level of primary nutrient ($s$), tissue level of waste ($w$), primary nutrient intake flux by healthy cells ($u$) and waste intake flux by healthy cells ($v$) as a function of $k$ for low (left column) and high (right column) values of $c$. The blue arrow in top-right plot marks the discontinuous transition highlighted in (**A**). The values of $c$ used in (**C**) are shown by yellow lines in (**A**).

In terms of the fraction of aberrant cells, instead, the thresholds (14) simply represent the values of $\phi$ that are consistent with the corresponding quasi-steady state nutrient level. So, for instance, the growth rate of aberrant cells is zero when $\phi \geq \phi_{gr}$, implying that $\phi_{gr}$ is an upper bound for the fraction of aberrant cells in the system. On the other hand $\phi_m$ is the maximum fraction of aberrant cells consistent with healthy cells satisfying their maintenance demands from the consumption of primary nutrient alone, while waste secretion occurs if $\phi < \phi_{sec}$. Therefore, a waste shuttle takes place when $\phi_m < \phi < \phi_{sec}$. The thresholds (14) are shown in Fig. 5B as functions of $k$ and $c$.

*Physical origin of the different phases.* Comparing the asymptotic aberrant fraction $\phi_f$ to the thresholds Eq. (14) yields three possible outcomes (namely $\phi_f \leq \phi_m$, $\phi_m < \phi_f < \phi_{sec}$ and $\phi_f \geq \phi_{sec}$). The corresponding regions in the ($k$, $c$) plane are shown in Fig. 5C. If $\phi_f \leq \phi_m$ (Region I in Fig. 5C), healthy cells satisfied their maintenance demand from $S$ alone throughout the dynamics, while aberrant cells exhibited sustained waste secretion. In this region, aberrant growth is limited by excess toxicity and aberrant take-over is only achieved if the turn-over rate $k$ is large enough to ensure remediation. A different scenario is obtained when $\phi_m < \phi_f < \phi_{sec}$ (Region II in Fig. 5C). Here, although the waste secreted by aberrant cells was initially uncoupled to the metabolism of healthy cells, eventually $\phi$ crosses the threshold $\phi_m$, a waste shuttle kicks-in and is sustained from then on. Again, the growth of aberrant cells is affected by waste toxicity, but its effects are now countered by a combination of healthy cell recycling and the turn-over rate $k$. Finally, if $\phi_f \geq \phi_{sec}$, waste secretion and recycling are not sustained because at some point in the dynamics $\phi$ becomes larger than the threshold $\phi_{sec}$. This leads to a state where aberrant growth can only be limited by nutrient availability. In this scenario, either the primary nutrient does not suffice to sustain a fully aberrant state ($\phi_f = \phi_{gr}$ in this case), or the aberrance eventually takes over the tissue ($\phi_f = 1$). In other words, $\phi_f = \min\{1, \phi_{gr}\}$. Additionally, $\phi_{sec}$ may become negative in Region III, defining a sub-region where waste is never secreted (Region IIIb in Fig. 5C). On the other hand, the portion of region III where $\phi_{sec} > 0$ (Region IIIa in Fig. 5C) has a transient waste secretion.

In Fig. 4A the ($k$, $c$) plane was split according to the role of waste. Confronting the two classifications one sees that region W+ coincides with region I as defined above, while region Sh. turns out to be composed of Regions II (where shuttling, once started, is sustained indefinitely) and IIIa (where waste secretion and shuttling are transient). Finally, region W− coincides with IIIb. The disjoint white areas of Fig. 4A where aberrant cells fail to expand to the entire tissue, one lying inside W− and the other inside W+, are now explicitly seen to trace back to different causes, namely nutrient scarcity for W− and waste toxicity for W+. Additionally, the sharp discontinuity highlighted by the blue boundary in Fig. 4A finds a simple interpretation. On the right side of the edge, aberrant growth is limited by toxicity and $\phi_f < \phi_{sec}$. On the left side, instead, waste is transient and nutrient availability becomes the determining factor, i.e. $\phi_f = \min\{1, \phi_{gr}\}$. The discontinuity therefore simply describes the transition from a state with $\phi_f < \phi_{sec}$ to one with $\phi_f = \min\{1, \phi_{gr}\}$.

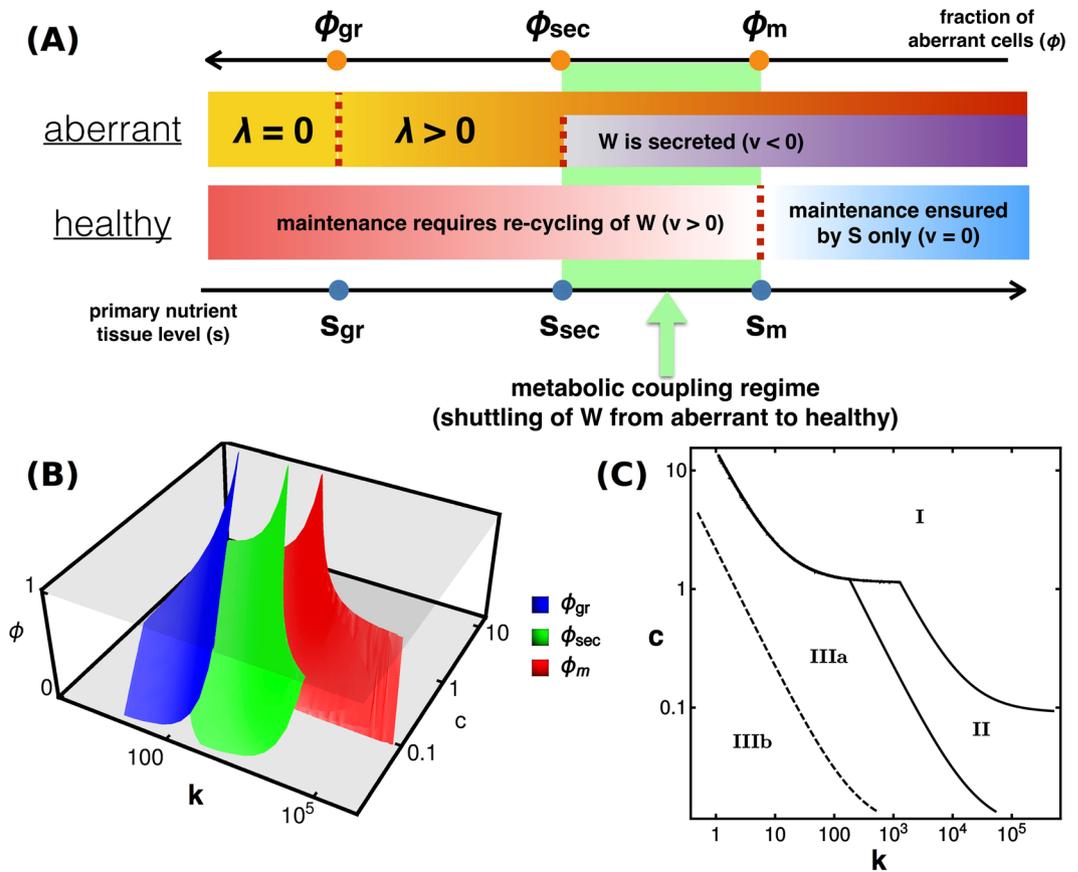

**Figure 5.** Summary of thresholds found in the spatially homogeneous model. (**A**) Aberrant cells can sustain a non-zero growth rate for $s > s_{gr}$ (or $\phi < \phi_{gr}$) and excrete waste for $s > s_{sec}$ (or $\phi < \phi_{sec}$). In turn, healthy cells need to re-cycle the waste in order to meet their maintenance goals when $s < s_m$ (or $\phi > \phi_m$). As a consequence, aberrant and healthy cells are coupled when $s_{sec} < s < s_m$ (or $\phi_m < \phi < \phi_{sec}$). Note that the $\phi$-axis (top) is reversed with respect to the $s$-axis (bottom). (**B**) $\phi_{gr}$, $\phi_{sec}$ and $\phi_m$ as functions of $k$ and $c$. (**C**) Phase structure in terms of the asymptotic aberrant fraction $\phi_f$. In Region I, $\phi_f \leq \phi_m$; in Region II, $\phi_m < \phi_f < \phi_{sec}$; in Region III, $\phi_f \geq \phi_{sec}$. The dashed line subdivides III according to whether $\phi_{sec} > 0$ (Region IIIa) or $\phi_{sec} \leq 0$ (Region IIIb). The phase labeled W+ (resp. W−) in the phase diagram shown in Fig. 4A coincides with Region I (resp. IIIb). Region Sh. is instead formed by Regions II and IIIa.

**The role of random fluctuations in the dynamics of the spatial model.** As pointed out above, while the phase diagrams in Figs 3 and 4 are qualitatively very similar, a major difference derives from fluctuations in the spatial model. In particular, in the parameter region where waste shuttling occurs, the spatial model exhibits a bistable behaviour such that random fluctuations may drive the tissue towards a fully aberrant or a fully healthy state (see Fig. 4A,B). Sample trajectories displaying this phenomenon are shown in Fig. 6. For the same initial conditions and parameters, we plot the fraction $\phi$ of aberrant cells versus time in a simulation of the mean-field model (blue line) and in several simulations of the spatial model (red and black lines). Trajectories where the spatial model's outcome qualitatively agrees with the mean-field model are shown in black, while cases in which the spatial model diverges from the mean-field behaviour are shown in red. One sees that fluctuations can bring the tissue to a healthy state when most trajectories converge to an aberrant one (as in Fig. 6A) or vice versa (as in Fig. 6B).

Intuition about how fluctuations may lead to different outcomes can be obtained by noting that, in terms of the thresholds derived for the mean-field model, toxicity limits aberrant growth only if $\phi_f < \phi_{sec}$ whereas waste accumulation ceases if the aberrant fraction reaches a value larger than $\phi_{sec}$. In between, a critical value of $\phi$ exists beyond which waste accumulation decreases if the fraction of aberrant cells increases and aberrant growth becomes positive. If, by random fluctuations, the aberrant cell population overcomes this threshold, it will then continue to grow up to the maximum value allowed by nutrient availability (i.e. $\min\{1, \phi_{gr}\}$). In such a case, the mean-field model fails to predict aberrant dominance. Likewise, if random evolution keeps $\phi$ below $\phi_{sec}$ long enough, toxicity caused by waste accumulation may lead to the disappearance of aberrant cells from the tissue.

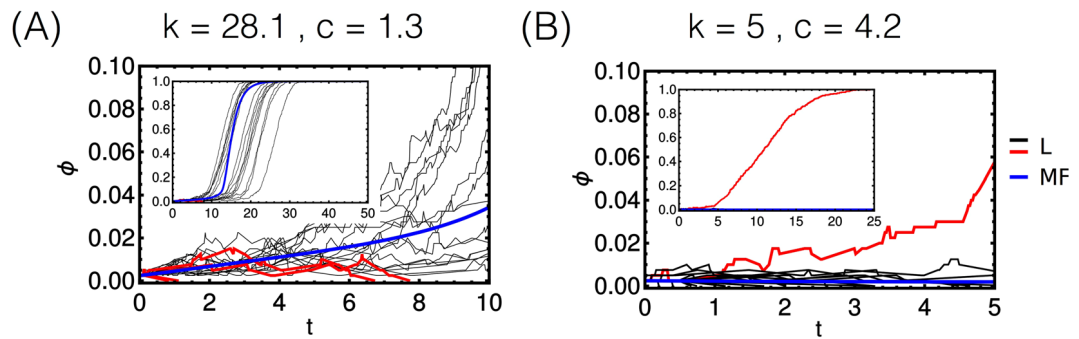

**Figure 6.** Bi-stability in the spatial model. (**A,B**) Aberrant cell fraction ($\phi$) plotted against time for the mean field model (blue line) and for several runs of the spatial model (black lines if the dynamical outcome of the spatial model agrees qualitatively with the spatially homogeneous model, red otherwise), for the different reported values of $k$ and $c$. Main panels: behaviour for short times. Insets: behaviour for ling times.

This clarifies that the equivalence between the mean-field and spatial models is not full. Remarkably, though, the mean qualitative behavior of the spatial model is well captured by its mean-field approximation.

Since the bistability is localized in a parameter region where waste recycling is prominent, we probed the link between the two occurrences. Specifically, we ran simulations of both models after disabling the shuttle by forcing $v_{he}=0$ for healthy cells. A comparison between the original and modified systems is showcased in Fig. 7. We zoom into the parameter region where disabling the shuttle has a stronger impact, corresponding to intermediate to high $c$ and low to intermediate $k$ (see Fig. 3). Random fluctuations resulted in bistability in the spatial model prominently in the region highlighted in blue in Fig. 7A, where aberrant growth is severely inhibited in absence of re-cycling by healthy cells (see Fig. 7B). This strongly suggests that the bistable stochastic dynamics of aberrant spread is enabled by the shuttling of overflow product by aberrant cells to healthy cells for recycling purposes. We then repeated this comparison in the homogeneous model, see Fig. 6C,D. Here silencing the shuttle only appears to bear a quantitative effect. As in Fig. 4A, we find a discontinuous edge (blue line) separating a region where growth is inhibited by toxicity from a region where growth is limited by nutrient availability. In absence of shuttling, the edge simply appears to shift towards slightly lower values of $c$ (see Fig. 7D), indicating that the recycling of toxic overflow products by healthy cells enables aberrant growth for larger supplies of primary nutrient.

## Discussion

Almost universally, cells sustaining high energy demands or fast growth undergo an overall shift in metabolic strategy that carries strong signatures in proteome composition and regulatory mechanisms[55–57]. Many explanations have been put forward for the seemingly inefficient use of nutrients that occurs as cells release partially metabolized substances into their surrounding environment[58]. In some cases, exceeding protein costs prevent the operation of high-yield respiratory at fast growth[13,59], but this scenario does not appear to be generic even for bacteria[60]. Likewise, crowding or capacity constraints account for the onset of the inefficient phase[10] but likely can't be invoked universally[61]. The challenge of understanding the fundamental trade-offs driving metabolic strategy selection in different cases is therefore still largely open. In turn, though, metabolic decisions by individual cells can determine significant population-level effects, for instance by inducing cross-feeding interactions[62], which then feed back into the fitness of individual cells.

Working along these lines, we have studied here how a spatially distributed population of interacting cells with very different energetic demands self-organizes in a shared environment coupled to a single nutrient reservoir. The model we define can be seen as a minimal approach to the analysis of the initial stages of the spread of an aberrance in a tissue. Many types of tumors generate an acidic microenvironment due to disregulated metabolism and lactate overflow[63]. Based on empirical evidence, we assume that (i) the tissue exchanges a primary nutrient with a reservoir that also serves as a sink for the overflow product, (ii) lactate shuttling may occur, i.e. non-aberrant neighbours of aberrant cells may intake lactate if the primary nutrient is insufficient to ensure their maintenance, (iii) the replication and death rate of aberrant cells are affected by the local level of lactate, and (iv) both the primary nutrient and lactate can diffuse across the tissue. From a physical viewpoint, the key assumption is that metabolism occurs on faster time scales than across-tissue diffusion, so that cells can effectively optimize their metabolic profile while local concentrations are fixed. Besides modeling details, this is a point where our work departs from the approaches presented e.g. in ref. 36, where a dynamic description of metabolism is employed. Our simplified (in this respect) scenario allows for the study of an extended, spatially organized system with a large number of cells.

Many physiological aspects are however highly schematized. For instance, our representation of the blood as a nutrient reservoir at fixed concentration ignores variability in blood glucose levels, which have been observed to be significant in specific conditions[64]. An increase of fluctuations is thought to be generically linked to the approach to dynamical transitions[65] and would impact the scenario discussed here. Likewise, we have employed a minimal representation of cellular metabolic networks that allows to focus on the organization of tissue-wide energy metabolism while keeping parameters to a controllable minimum. Performing a similar analysis starting from genome-scale models of metabolism would require a significant additional computational effort focused on selecting thermodynamically viable flux patterns[66, 67]. The qualitative scenario emerging from the present study however is unlikely to be altered in a more complex, parameter-rich setting.

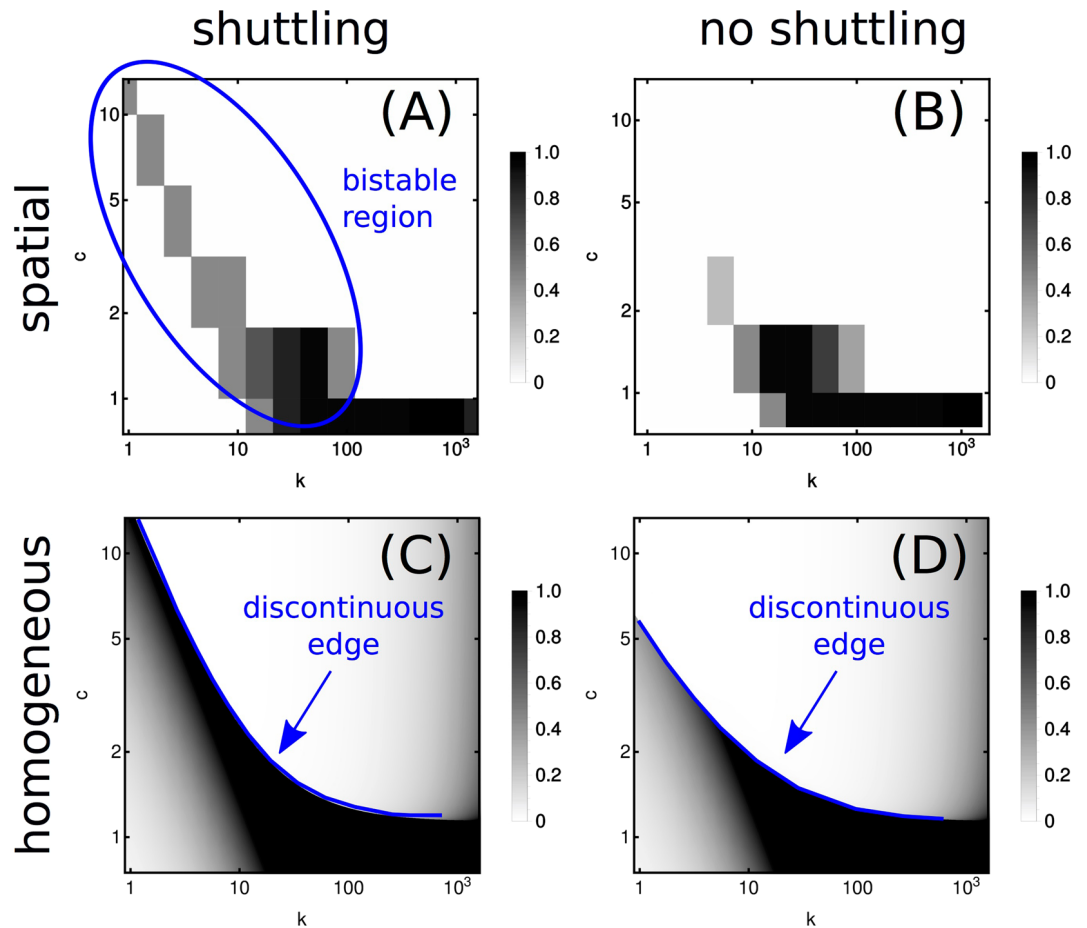

**Figure 7.** Effect of shuttling in spatial and homogeneous model. Phase portraits obtained for the spatial (**A**,**B**) and the homogeneous (**C**,**D**) models, with enabled (**A**,**C**) and disabled (**B**,**D**) overflow product shuttling. Notice that the range of values of *c* and *k* are narrower with respect to those employed in Figs 3 and 4A. Outside the shown ranges, turning off re-cycling by healthy cells appears to bear no impact on the final state of the tissue. Notice that aberrant growth in the region of bistability shown in (**A**) is significantly reduced by silencing the shuttle, see (**B**).

Our results are summed up in the phase portrait of Fig. 3 (similar to the outcome diagram obtained for cross-feeding microbial populations in ref. 36), where the asymptotic state of the tissue (in terms of the fraction $\phi$ of aberrant cells) is shown as a function of the primary nutrient level *s* and of the tissue-reservoir turn-over rate *k*. Three phases appear. In the phase labeled W− (sufficiently small *c* and *k*), the spreading dynamics of the Warburg-like phenotype is limited by nutrient availability and no shuttling of overflow product occurs. Upon increasing *c* and/or *k* one enters the Sh. region. All over this phase, the aberrant spreading dynamics is sustained by healthy cells re-cycling harmful overflow products. However, the long-term state of the tissue is typically healthy for large *c* and small *k* (in which case waste toxicity limits aberrant take-over in spite of microenvironmental cooperation) and aberrant for large *k* and small *c* (when a combination of efficient waste re-cycling and fast turn-over with the blood leads to a fully aberrant state). Finally, region W+ is found at large enough *c* and *k*. Here, no re-cycling takes place and the long-term state of the tissue is determined essentially by its coupling to the blood (healthy for slow turn-over, aberrant for fast turn-over). We have tracked the physical origin of these phases through a simplified, mathematically tractable version of the model whose solution ultimately relies on just 3 variables (the fraction of aberrant cells and the levels of primary nutrient and waste) and 3 (combinations of) parameters given by the thresholds in (13) or (14). The main feature escaping the solvable model is the presence of bistability, occurring in the spatial model for large *c* and small *k*. Here, the final state of the tissue can be strongly influenced by random fluctuations occurring during the dynamics. This phenomenon appears to be coupled to the shuttling of overflow products between aberrant and healthy cells, as revealed by disabling re-cycling by non-aberrant cells in Fig. 7.

Stretching out the information encoded in the phase structure, our simulations suggest different qualitative strategies to slow down aberrant growth. At intermediate *k*, one may either deprive aberrant cells of primary nutrients by decreasing *c* or foster growth suppression via tissue toxification by increasing *c*. For higher *k* and high enough *c*, the aberrance spread can be halted by toxicity. In this regime increasing *c* is the only way to lead the tissue back to a healthy state. Increased glucose supply in tumors is known to lead to faster acidification[42], an effect that is usually connected to the production of lactate. Moreover, glucose infusion leads to less efficient blood-tissue turn-over, probably due to enhanced viscosity[68]. This aspect also contributes to acidification[42]. Notice that arterial vasodilators may selectively increase perfusion in tissues, leading to a fall in their pH[69]. Our results also suggest that targeting lactate shuttling in tumors[42, 70] may be effective in driving a tissue from an aberrant state towards a healthy state, provided *k* is sufficiently low and *c* is high, in agreement with experimental findings[29, 71]. It is however important to remark that the role of lactate in cancer is turning out to be much broader than the aspects on which we focus here, especially in connection to its interaction with the immune system[72]. In this sense, our model is more relevant to the understanding of early lesions consisting of a few aberrant cells, which are effectively invisible to the immune system due to their small size.

## Methods

### Choice of parameters.
Our choice of parameters was guided by the example of the glucose-lactate pair (for primary nutrient and overflow product, respectively) in cancer growth.

The ATP yields appearing in (1) were chosen following the simple glycolytic models of refs [28], [50], i.e. $Y_u = 1$ and $Y_r = 19$ (both giving moles of ATP output per mole of *S* imported). The ATP demand related to maintenance for mammalian cells is usually reported to be around[73] $e_m = 10$ pmol/day/$h^3$, where $h \simeq 0.01$ mm denotes the typical linear dimension of a cell (see also below). The parameters $\lambda_{max}$ and $K_E$ in (6) were fitted to agree with the approximate linear relation between $e - e_m$ and the growth rate at low energy production rates (with a proportionality constant close to 10 pmol/cell)[73] and with the observed maximum growth rate of mammalian cells (close to 3 day$^{-1}$, see ref. [74]), obtaining $\lambda_{max} = 2$ day$^{-1}$ and $K_E = 20$ pmol/day/$h^3$.

The most commonly expressed glucose transporter in mammalian cells is GLUT1, with a reported affinity of 1 mM[75, 76]. The average physiological concentration of glucose in the blood is around 5 mM. Assuming that the uptake of healthy cells follows a Michaelis-Menten law and that this concentration is enough to sustain the maintenance demand, we can solve for $u_{max}$ in the equation $u_m = u_{MM}(s)$ to obtain a value of $u_{max} = 0.5$ pmol/day/$h^3$. We re-defined the glucose to lactate stoichiometric ratio to one by doubling glucose levels and setting $K_s = 2$ mM, $u_{max} = 1$ pmol/day/$h^3$. To model the up-regulation of glucose transporters in growth-optimizing aberrant cells we use $\alpha = 100$ in (4). With the example of lactate in mind, with a measured Michaelis constant ranging between 1 mM and 10 mM, we set $K_w = 5$ mM, while the maximum uptake of this metabolite has been reported[77–79] around $v_{max} = 100$ pmol/day/$h^3$.

Measured values of the glucose diffusion coefficient in tissue span values between $10^{-6}$ cm$^2$/sec and $10^{-5}$ cm$^2$/sec, depending on the tissue and measurement technique[80–82]. We adopted an intermediate common value of the diffusion coefficient for substrate and waste, of $1.5 \times 10^{-6}$ cm$^2$/sec. Cells in the 2-dimensional grid were assumed to be a distance *h* apart that reflects the number of cells per unit area in flat tissues. This value varies widely across different tissues[83, 84], and we set it to $10^4$ cells/mm$^2$, which implies a typical distance *h* between cells around 0.01 mm. Coupling this to the coefficients of the discretized Laplacian operator in the 2-dimensional grid ($d/h^2$), we obtain the value 0.06 sec$^{-1}$ for the effective diffusion constants $D_S$ and $D_W$.

The inhibitory effects of waste on growth are again modeled after lactate. Quantitative expressions have been obtained from *in vitro* studies. We use the values $K_{inh} = 8$ mM, $K_{tox} = 15$ mM and $\sigma_{max} = 2$ day$^{-1}$ from ref. [53], but see also refs [85–87].

We assume that cells have a maximum capacity to fully oxidize the precursor *P* arising e.g. from a limitation in the proteome fraction available for enzyme production[12] or by limited mitochondrial efficiency[15]. Following[28, 50], we set $r_{max} = 6.48$ pmol/day/cell.

Finally, metabolite levels are measured in units of $K_{tox}$ (so that $K_{tox} = 1$), time is measured in units of $1/\sigma_{max}$, and fluxes are re-scaled by $K_{tox}\sigma_{max}$. The detailed definition of non-dimensional parameters is provided in the Supporting Text.

Note that, with these choices, the standard values of *c* ranging from hypoglycaemic to hyperglycaemic conditions covers about an order of magnitude in the central part of the vertical axis of Figs 3 and 4A. Therefore the overall range of values for *c* and *k* over which we characterize both the spatial model and the simplified one is considerably broader. Note however that the physiological range contains each of the different phases and sub-phases that we characterize.

### Simulation details.
Numerical simulations of the lattice model were carried out in Julia[88]. The script is available from Github (https://github.com/cossio/aberr). To speed up simulations, we exploited the fact that the spatial distribution of metabolite levels equilibrates fast compared to cell dynamics. Specifically, at each step, we checked whether metabolite levels have equilibrated, and advanced cell dynamics to the next division/death event (following the Gillespie algorithm[89]) whenever an equilibrium had been achieved. The asymptotic fraction of aberrant cells in the mean field model was computed by simulating the model's ODEs for a very long time using the 'NSolve' routine from Mathematica.

## References

1. Wolfe, A. J. The acetate switch. *Microbiology and Molecular Biology Reviews* **69**, 12–50 (2005).
2. De Deken, R. The Crabtree effect: a regulatory system in yeast. *Microbiology* **44**, 149–156 (1966).
3. Westerblad, H., Bruton, J. D. & Katz, A. Skeletal muscle: energy metabolism, fiber types, fatigue and adaptability. *Experimental Cell Research* **316**, 3093–3099 (2010).
4. Vander Heiden, M. G., Cantley, L. C. & Thompson, C. B. Understanding the Warburg effect: the metabolic requirements of cell proliferation. *Science* **324**, 1029–1033 (2009).

5. Pearce, E. L., Poffenberger, M. C., Chang, C.-H. & Jones, R. G. Fueling immunity: insights into metabolism and lymphocyte function. *Science* **342**, 1242454 (2013).
6. Liberti, M. V. & Locasale, J. W. The Warburg effect: how does it benefit cancer cells? *Trends in Biochemical Sciences* **41**, 211–218 (2016).
7. Shlomi, T., Benyamini, T., Gottlieb, E., Sharan, R. & Ruppin, E. Genome-scale metabolic modeling elucidates the role of proliferative adaptation in causing the Warburg effect. *PLoS Computational Biology* **7**, e1002018 (2011).
8. De Martino, D., Capuani, F. & De Martino, A. Inferring metabolic phenotypes from the exometabolome through a thermodynamic variational principle. *New Journal of Physics* **16**, 115018 (2014).
9. Dai, Z., Shestov, A. A., Lai, L. & Locasale, J. W. A flux balance of glucose metabolism clarifies the requirements of the Warburg effect. *Biophysical Journal* **111**, 1088–1100 (2016).
10. Vazquez, A. & Oltvai, Z. N. Macromolecular crowding explains overflow metabolism in cells. *Scientific Reports* **6**, 31007 (2016).
11. O'Brien, E. J., Lerman, J. A., Chang, R. L., Hyduke, D. R. & Palsson, B. Ø. Genome-scale models of metabolism and gene expression extend and refine growth phenotype prediction. *Molecular Systems Biology* **9**, 693 (2013).
12. Basan, M. *et al.* Overflow metabolism in Escherichia coli results from efficient proteome allocation. *Nature* **528**, 99–104 (2015).
13. Mori, M., Hwa, T., Martin, O. C., De Martino, A. & Marinari, E. Constrained allocation flux balance analysis. *PLoS Computational Biology* **12**, e1004913 (2016).
14. Zhuang, K., Vemuri, G. N. & Mahadevan, R. Economics of membrane occupancy and respiro-fermentation. *Molecular Systems Biology* **7**, 500 (2011).
15. Vazquez, A. Limits of aerobic metabolism in cancer cells. *bioRxiv* 020461 (2015).
16. Zhou, W. *et al.* HIF1$\alpha$ induced switch from bivalent to exclusively glycolytic metabolism during ESC-to-EpiSC/hESC transition. *EMBO Journal* **31**, 2103–2116 (2012).
17. Huberts, D. H., Niebel, B. & Heinemann, M. A flux-sensing mechanism could regulate the switch between respiration and fermentation. *FEMS Yeast Research* **12**, 118–128 (2012).
18. Stark, H. *et al.* Causes of upregulation of glycolysis in lymphocytes upon stimulation. A comparison with other cell types. *Biochimie* **118**, 185–194 (2015).
19. Schuster, S. *et al.* Mathematical models for explaining the Warburg effect: a review focussed on ATP and biomass production. *Biochemical Society Transactions* **43**, 1187–1194 (2015).
20. Pfeiffer, T., Schuster, S. & Bonhoeffer, S. Cooperation and competition in the evolution of ATP-producing pathways. *Science* **292**, 504–507 (2001).
21. Pfeiffer, T. & Bonhoeffer, S. Evolutionary consequences of tradeoffs between yield and rate of ATP production. *Zeitschrift für Physikalische Chemie* **216**, 51 (2002).
22. Brooks, G. A. Cell–cell and intracellular lactate shuttles. *Journal of Physiology* **587**, 5591–5600 (2009).
23. Koukourakis, M. I., Giatromanolaki, A., Harris, A. L. & Sivridis, E. Comparison of metabolic pathways between cancer cells and stromal cells in colorectal carcinomas: a metabolic survival role for tumor-associated stroma. *Cancer Research* **66**, 632–637 (2006).
24. Brooks, G. A. Intra- and extra-cellular lactate shuttles. *Medicine and Science in Sports and Exercise* **32**, 790–799 (2000).
25. Bélanger, M., Allaman, I. & Magistretti, P. J. Brain energy metabolism: focus on astrocyte-neuron metabolic cooperation. *Cell Metabolism* **14**, 724–738 (2011).
26. Massucci, F. A. *et al.* Energy metabolism and glutamate-glutamine cycle in the brain: a stoichiometric modeling perspective. *BMC Systems Biology* **7**, 103 (2013).
27. Kianercy, A., Veltri, R. & Pienta, K. J. Critical transitions in a game theoretic model of tumour metabolism. *Interface Focus* **4**, 20140014 (2014).
28. Capuani, F., De Martino, D., Marinari, E. & De Martino, A. Quantitative constraint-based computational model of tumor-to-stroma coupling via lactate shuttle. *Scientific Reports* **5** (2015).
29. Sonveaux, P. *et al.* Targeting lactate-fueled respiration selectively kills hypoxic tumor cells in mice. *Journal of Clinical Investigation* **118**, 3930–3942 (2008).
30. Hay, N. Reprogramming glucose metabolism in cancer: can it be exploited for cancer therapy? *Nature Reviews Cancer* (2016).
31. Pavlides, S. *et al.* The reverse Warburg effect: aerobic glycolysis in cancer associated fibroblasts and the tumor stroma. *Cell Cycle* **8**, 3984–4001 (2009).
32. Pietras, K. & Östman, A. Hallmarks of cancer: interactions with the tumor stroma. *Experimental Cell Research* **316**, 1324–1331 (2010).
33. Sousa, C. M. *et al.* Pancreatic stellate cells support tumour metabolism through autophagic alanine secretion. *Nature* **536**, 479–483 (2016).
34. MacLean, R. C. & Gudelj, I. Resource competition and social conflict in experimental populations of yeast. *Nature* **441**, 498–501 (2006).
35. MacLean, R. C. *et al.* A mixture of "cheats" and "co-operators" can enable maximal group benefit. *PLoS Biology* **8**, e1000486 (2010).
36. Gudelj, I. *et al.* Stability of cross-feeding polymorphisms in microbial communities. *PLoS Computational Biology* **12**, e1005269 (2016).
37. Ozturk, S. S., Riley, M. R. & Palsson, B. O. Effects of ammonia and lactate on hybridoma growth, metabolism, and antibody production. *Biotechnology and Bioengineering* **39**, 418–431 (1992).
38. Schneider, M., Marison, I. W. & von Stockar, U. The importance of ammonia in mammalian cell culture. *Journal of Biotechnology* **46**, 161–185 (1996).
39. Gatenby, R. A. & Gawlinski, E. T. A reaction-diffusion model of cancer invasion. *Cancer Research* **56**, 5745–5753 (1996).
40. Gatenby, R. A. & Gillies, R. J. Why do cancers have high aerobic glycolysis? *Nature Reviews Cancer* **4**, 891–899 (2004).
41. Gatenby, R. A., Gawlinski, E. T., Gmitro, A. F., Kaylor, B. & Gillies, R. J. Acid-mediated tumor invasion: A multidisciplinary study. *Cancer Research* **66**, 5216–5223 (2006).
42. Tannock, I. F. & Rotin, D. Acid pH in tumors and its potential for therapeutic exploitation. *Cancer Research* **49**, 4373–4384 (1989).
43. Roe, A. J., O'Byrne, C., McLaggan, D. & Booth, I. R. Inhibition of Escherichia coli growth by acetic acid: a problem with methionine biosynthesis and homocysteine toxicity. *Microbiology* **148**, 2215–2222 (2002).
44. Casey, G. P. & Ingledew, W. M. Ethanol tolerance in yeasts. *CRC Critical Reviews in Microbiology* (2008).
45. Young, J. D. Metabolic flux rewiring in mammalian cell cultures. *Current Opinion in Biotechnology* **24**, 1108–1115 (2013).
46. Martnez, V. S. *et al.* Flux balance analysis of CHO cells before and after a metabolic switch from lactate production to consumption. *Biotechnology and Bioengineering* **110**, 660–666 (2013).
47. Ferreira, S. Jr., Martins, M. & Vilela, M. Reaction-diffusion model for the growth of avascular tumor. *Physical Review E* **65**, 021907 (2002).
48. Jiang, Y., Pjesivac-Grbovic, J., Cantrell, C. & Freyer, J. P. A multiscale model for avascular tumor growth. *Biophysical Journal* **89**, 3884–3894 (2005).
49. Roose, T., Chapman, S. J. & Maini, P. K. Mathematical models of avascular tumor growth. *SIAM Review* **49**, 179–208 (2007).
50. Vazquez, A., Liu, J., Zhou, Y. & Oltvai, Z. N. Catabolic efficiency of aerobic glycolysis: the Warburg effect revisited. *BMC Systems Biology* **4**, 1 (2010).
51. Beg, Q. K. *et al.* Intracellular crowding defines the mode and sequence of substrate uptake by Escherichia coli and constrains its metabolic activity. *Proceedings of the National Academy of Sciences* **104**, 12663–12668 (2007).

52. Liu, Y. A simple thermodynamic approach for derivation of a general Monod equation for microbial growth. *Biochemical Engineering Journal* **31**, 102–105 (2006).
53. Bree, M. A., Dhurjati, P., Geoghegan, R. F. & Robnett, B. Kinetic modelling of hybridoma cell growth and immunoglobulin production in a large-scale suspension culture. *Biotechnology and Bioengineering* **32**, 1067–1072 (1988).
54. Bertolazzi, E. A combination formula of Michaelis-Menten-Monod type. *Computers and Mathematics with Applications* **50**, 201–215 (2005).
55. Molenaar, D., van Berlo, R., de Ridder, D. & Teusink, B. Shifts in growth strategies reflect tradeoffs in cellular economics. *Molecular Systems Biology* **5**, 323 (2009).
56. Hui, S. *et al.* Quantitative proteomic analysis reveals a simple strategy of global resource allocation in bacteria. *Molecular Systems Biology* **11**, 784 (2015).
57. Borkowski, O. *et al.* Translation elicits a growth rate-dependent, genome-wide, differential protein production in bacillus subtilis. *Molecular Systems Biology* **12**, 870 (2016).
58. Diaz-Ruiz, R., Rigoulet, M. & Devin, A. The Warburg and Crabtree effects: on the origin of cancer cell energy metabolism and of yeast glucose repression. *Biochimica et Biophysica Acta (BBA)-Bioenergetics* **1807**, 568–576 (2011).
59. Flamholz, A., Noor, E., Bar-Even, A., Liebermeister, W. & Milo, R. Glycolytic strategy as a tradeoff between energy yield and protein cost. *Proceedings of the National Academy of Sciences* **110**, 10039–10044 (2013).
60. Goel, A. *et al.* Protein costs do not explain evolution of metabolic strategies and regulation of ribosomal content: does protein investment explain an anaerobic bacterial Crabtree effect? *Molecular Microbiology* **97**, 77–92 (2015).
61. Basan, M. *et al.* Inflating bacterial cells by increased protein synthesis. *Molecular Systems Biology* **11**, 836 (2015).
62. Hummert, S. *et al.* Evolutionary game theory: cells as players. *Molecular BioSystems* **10**, 3044–3065 (2014).
63. Hsu, P. P. & Sabatini, D. M. Cancer cell metabolism: Warburg and beyond. *Cell* **134**, 703–707 (2008).
64. Pisarchik, A. N., Pochepen, O. N. & Pisarchyk, L. A. Increasing blood glucose variability is a precursor of sepsis and mortality in burned patients. *PloS One* **7**, e46582 (2012).
65. Challet, D., De Martino, A. & Marsili, M. Dynamical instabilities in a simple minority game with discounting. *Journal of Statistical Mechanics: Theory and Experiment* **2008**, L04004 (2008).
66. De Martino, D., Capuani, F., Mori, M., De Martino, A. & Marinari, E. Counting and correcting thermodynamically infeasible flux cycles in genome-scale metabolic networks. *Metabolites* **3**, 946–966 (2013).
67. Desouki, A. A., Jarre, F., Gelius-Dietrich, G. & Lercher, M. J. Cyclefreeflux: efficient removal of thermodynamically infeasible loops from flux distributions. *Bioinformatics* btv096 (2015).
68. Sevick, E. M. & Jain, R. K. Blood flow and venous pH of tissue-isolated Walker 256 carcinoma during hyperglycemia. *Cancer Research* **48**, 1201–1207 (1988).
69. Okunieff, P., Kallinowski, F., Vaupel, P. & Neuringer, L. J. Effects of hydralazine-induced vasodilation on the energy metabolism of murine tumors studied by *in vivo* 31p-nuclear magnetic resonance spectroscopy. *Journal of the National Cancer Institute* **80**, 745–750 (1988).
70. Gerweck, L. E. & Seetharaman, K. Cellular pH gradient in tumor versus normal tissue: potential exploitation for the treatment of cancer. *Cancer Research* **56**, 1194–1198 (1996).
71. Pisarsky, L. *et al.* Targeting metabolic symbiosis to overcome resistance to anti-angiogenic therapy. *Cell Reports* **15**, 1161–1174 (2016).
72. Brand, A. *et al.* LDHA-associated lactic acid production blunts tumor immunosurveillance by T and NK cells. *Cell Metabolism* **24**, 657–671 (2016).
73. Kilburn, D., Lilly, M. & Webb, F. The energetics of mammalian cell growth. *Journal of Cell Science* **4**, 645–654 (1969).
74. Kiparissides, A., Koutinas, M., Kontoravdi, C., Mantalaris, A. & Pistikopoulos, E. N. 'Closing the loop' in biological systems modeling – from the in silico to the *in vitro*. *Automatica* **47**, 1147–1155 (2011).
75. Gaertner, J. G. & Dhurjati, P. Fractional factorial study of hybridoma behavior. 2. Kinetics of nutrient uptake and waste production. *Biotechnology Progress* **9**, 309–316 (1993).
76. Zhao, F.-Q. & Keating, A. F. Functional properties and genomics of glucose transporters. *Current Genomics* **8**, 113–128 (2007).
77. Jackson, V. N. & Halestrap, A. P. The kinetics, substrate, and inhibitor specificity of the monocarboxylate (lactate) transporter of rat liver cells determined using the fluorescent intracellular pH indicator, 2′,7′-bis(carboxyethyl)-5 (6)-carboxyfluorescein. *Journal of Biological Chemistry* **271**, 861–868 (1996).
78. Tildon, J. T., McKenna, M. C., Stevenson, J. & Couto, R. Transport of L-lactate by cultured rat brain astrocytes. *Neurochemical Research* **18**, 177–184 (1993).
79. Spencer, T. L. & Lehninger, A. L. L-lactate transport in Ehrlich ascites-tumour cells. *Biochemical Journal* **154**, 405–414 (1976).
80. Groebe, K., Erz, S. & Mueller-Klieser, W. Glucose diffusion coefficients determined from concentration profiles in EMT6 tumor spheroids incubated in radioactively labeled L-glucose. In *Oxygen Transport to Tissue XVI*, 619–625 (Springer, 1994).
81. Bashkatov, A. N. *et al.* Estimation of glucose diffusion coefficient in scleral tissue. In *Saratov Fall Meeting' 99*, 345–355 (International Society for Optics and Photonics, 2000).
82. Ghosn, M. G., Tuchin, V. V. & Larin, K. V. Depth-resolved monitoring of glucose diffusion in tissues by using optical coherence tomography. *Optics Letters* **31**, 2314–2316 (2006).
83. Radisic, M. *et al.* Oxygen gradients correlate with cell density and cell viability in engineered cardiac tissue. *Biotechnology and Bioengineering* **93**, 332–343 (2006).
84. Chan, E. H., Chen, L., Rao, J. Y., Yu, F. & Deng, S. X. Limbal basal cell density decreases in limbal stem cell deficiency. *American Journal of Ophthalmology* **160**, 678–684 (2015).
85. Omasa, T., Higashiyama, K.-I., Shioya, S. & Suga, K.-I. Effects of lactate concentration on hybridoma culture in lactate-controlled fed-batch operation. *Biotechnology and Bioengineering* **39**, 556–564 (1992).
86. Batt, B. C. & Kompala, D. S. A structured kinetic modeling framework for the dynamics of hybridoma growth and monoclonal antibody production in continuous suspension cultures. *Biotechnology and Bioengineering* **34**, 515–531 (1989).
87. Pörtner, R. & Schäfer, T. Modelling hybridoma cell growth and metabolism: a comparison of selected models and data. *Journal of Biotechnology* **49**, 119–135 (1996).
88. Bezanson, J., Edelman, A., Karpinski, S. & Shah, V. B. Julia: A fresh approach to numerical computing. *arXiv:1411.1607* (2014).
89. Gillespie, D. T. Stochastic simulation of chemical kinetics. *Annual Review of Physical Chemistry* **58**, 35–55 (2007).

## Acknowledgements

Work supported by the Italian Institute of Technology's Seed Project DREAM and by the European Union's Horizon 2020 research and innovation programme MSCA-RISE-2016 under grant agreement No 734439 INFERNET.

## Author Contributions

A.D.M. and R.M. conceived research; all authors designed research; J.F.C.D. performed simulations; all authors analyzed results; all authors wrote the manuscript.

**Additional Information**

**Supplementary information** accompanies this paper at doi:10.1038/s41598-017-03342-3

**Competing Interests:** The authors declare that they have no competing interests.

**Publisher's note:** Springer Nature remains neutral with regard to jurisdictional claims in published maps and institutional affiliations.

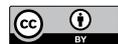 **Open Access** This article is licensed under a Creative Commons Attribution 4.0 International License, which permits use, sharing, adaptation, distribution and reproduction in any medium or format, as long as you give appropriate credit to the original author(s) and the source, provide a link to the Creative Commons license, and indicate if changes were made. The images or other third party material in this article are included in the article's Creative Commons license, unless indicated otherwise in a credit line to the material. If material is not included in the article's Creative Commons license and your intended use is not permitted by statutory regulation or exceeds the permitted use, you will need to obtain permission directly from the copyright holder. To view a copy of this license, visit http://creativecommons.org/licenses/by/4.0/.

© The Author(s) 2017